\documentclass[11pt,leqno]{article}

\usepackage{bm}
\usepackage{epsfig,graphics}
\usepackage{amsfonts}

\usepackage{amssymb}
\newcommand{\nc}{\newcommand}
\nc{\dps}{\displaystyle}

\newtheorem{theorem}{Theorem}

\nc{\RR}{\mbox{\rm I$\!$R}}

 \newcommand{\beqn}{\begin{eqnarray}}
 \newcommand{\eeqn}{\end{eqnarray}}
 \newcommand{\be}{\begin{equation}}
 \newcommand{\ee}{\end{equation}}
 \newcommand{\ba}{\begin{array}}
 \newcommand{\ea}{\end{array}}
 
 \newcommand{\pa}{\partial}
 \newcommand{\re}{\ref}
 \newcommand{\ci}{\cite}
 \newcommand{\ds}{\displaystyle}
 \newcommand{\la}{\label}
 \newcommand{\bfr}{\begin{flushright}}
 \newcommand{\efr}{\end{flushright}}
 
 \newcommand{\rIm}{{\rm Im\5}}
 \newcommand{\rRe}{{\rm Re\5}}
\newcommand{\bfl}{\begin{flushleft}}
\newcommand{\efl}{\end{flushleft}}
\newcommand{\fr}{\frac}

\newcommand{\ov}{\overline}
\newcommand{\st}{\stackrel}

\newcommand{\ti}{\tilde}

\newcommand{\n}{{\bf n}}

\newcommand{\e}{{\bf e}}

\newcommand{\x}{{\bf x}}
\newcommand{\y}{{\bf y}}

\newcommand{\bk}{{\bf k}}

\newcommand{\cm}{{\rm m}}

\newcommand{\bA}{{\bf A}}

\newcommand{\cO}{{\cal O}}

\newcommand{\bj}{{\bf j}}

\newcommand{\ka}{\kappa}
\newcommand{\ve}{\varepsilon}
\newcommand{\vp}{\varphi}

\newcommand{\De}{\Delta}

\newcommand{\om}{\omega}

\newcommand{\na}{\nabla}

\newcommand{\lam}{\lambda}

 \newcommand{\h}{{h^{\hspace{-2.5mm}-}}}

\newcommand{\5}{{\hspace{0.5mm}}}

\newcommand\R{{\mathbb R}}

\newcommand\nab{{\bm \na}}

\newtheorem{qtheorem}{QTheorem}[section]

\newtheorem{defin}[theorem]{Definition}

\newtheorem{lemma}[theorem]{Lemma}
\newtheorem{example}[theorem]{Example}
\newtheorem{exercice}[theorem]{Exercise}
\newtheorem{remark}[theorem]{Remark}
\newtheorem{remarks}[theorem]{Remarks}
\newtheorem{cor}[theorem]{Corollary}
\newtheorem{pro}[theorem]{Proposition}

\newtheorem{coms}[theorem]{Comments}

\newcommand{\bd}{\begin{defin}}
 \newcommand{\ed}{\end{defin}}
\newcommand{\bt}{\begin{theorem}}
 \newcommand{\et}{\end{theorem}}
\newcommand{\bqt}{\begin{qtheorem}}
 \newcommand{\eqt}{\end{qtheorem}}

\newcommand{\bp}{\begin{pro}}
 \newcommand{\ep}{\end{pro}}

\newcommand{\bl}{\begin{lemma}}
 \newcommand{\el}{\end{lemma}}
\newcommand{\bc}{\begin{cor}}
 \newcommand{\ec}{\end{cor}}

\newcommand{\bex}{\begin{example}}
 \newcommand{\eex}{\end{example}}
\newcommand{\bexs}{\begin{examples}}
 \newcommand{\eexs}{\end{examples}}

\newcommand{\bexe}{\begin{exercice}}
 \newcommand{\eexe}{\end{exercice}}

\newcommand{\br}{\begin{remark} }
 \newcommand{\er}{\end{remark}}
\newcommand{\brs}{\begin{remarks}}
 \newcommand{\ers}{\end{remarks}}

\newcommand{\bcoms}{\begin{coms}}
\newcommand{\ecoms}{\end{coms}}

\begin{document}


\begin{center}
~
\\
{\Huge\bf On wave theory of the photoeffect}
\\~
\\~
\\
\vspace{15mm}
{\large A.I.Komech
\footnote{
Supported partly by
Mechanical-Mathematical department of Moscow State University 
(M.V.Lomonosov),
Alexander von Humboldt Research Award, and
 by the Austrian Science Fund (FWF): P22198-N13.}}\\
{\it
Faculty of Mathematics of Vienna University and\\
Institute for Information Transmission Problems\\
of Russian Academy of Sciences\\
}
\bigskip\bigskip\bigskip\bigskip\bigskip
{\Large\bf Abstract}
\end{center}
\vspace{-0mm}

The survey of the theory of the photoelectric effect 
is given. We start with Lenard's empirical observations 
and their phenomenological explanation by Einstein.
Further we present the updated version of 
 Wentzel's first order 
perturbation theory  of the photoeffect. 
Our main goal is a justification of the Wentzel theory 
with the limiting amplitude principle.
The corresponding nonlinear theory, relying on the 
Maxwell-Schr\"odinger coupled equations, is still open problem.



\newpage
\setcounter{subsection}{0}
\setcounter{theorem}{0}
\setcounter{equation}{0}
\section{Introduction}\la{sPE}

The {\it photoelectric effect} has been discovered by 
H. Hertz, and studied experimentally by P. Lenard.
First theoretical explanation has been done by A. Einstein 
who suggested the corpuscular theory of light 
introducing the ``photons'' which are particles of the light.
In the framework of 
the Schr\"odinger theory the effect has been 
described first by G. Wentzel who calculated the angular
distribution of the photocurrent.
The calculation  relies on the 
perturbation procedure 
applied to the coupled Maxwell-Schr\"odinger equations.

The  photoeffect
 first was observed by H. Hertz in 1887:
he discovered the discharge of negatively charged {\it electroscope}
under the electromagnetic radiation of 
very short wavelength, such as visible or ultraviolet light.
This discharge was treated as
an emission of the 
 electrons from metals
due to their absorption of energy from electromagnetic radiation.

In 1902
P. Lenard systematically studied the behavior of the ``photoelectrons'',
i.e.
emitted electrons, in external  electric and magnetic fields.
His conclusions 
were the following:
\medskip\\
{\bf L1.} The saturation photocurrent is proportional to intensity of 
incident  light.
\medskip\\
{\bf L2.} The photocurrent is not zero only
for sufficiently small wavelength, i.e. for 
high frequencies: 
\be\la{redbound}
|\om|>\om_{\rm red}~,
\ee
where
$\om_{\rm red}$ is called the {\it red bound} of the 
photoelectric effect which depends on the substance
but {\it does not depend on the  intensity of 
the  light}.
\medskip\\
{\bf L3.} The photocurrent vanishes 
if the 
stopping voltage $U_{\rm stop}$
is applied; the minimal $U_{\rm stop}$
also  depends on the substance
but {\it does not depend on the  intensity of 
the  light}. Moreover,  the minimal $U_{\rm stop}$
increases  for decreasing wave length of the incident light. 
\medskip

This independence of $\om_{\rm red}$ and minimal $U_{\rm stop}$
of the intensity of   light was the
main difficulty in theoretical explanation of the Lenard observations.
This independence seemed to constitute a new misterious phenomenon, 
which newer occured in  classical physics.
\medskip

In 1905, Einstein proposed a revolutionary interpretation, by 
developing the Planck's discretization for 
the energy of the Maxwell field oscillators with steps
$\h\om$.
Namely, he suggested that the matter absorbs the
light energy also by the discrete portions $\h\om$.
This corresponds to the treatment of light 
with frequency $\om$
as a beam of particles, 
called ``photons'', with energy $\h\om$. 
{\it Einstein's rules}
 for the  photoelectric effect are the following:
\medskip\\
{\bf E1.} The flux of photons is proportional to the intensity 
of the incident light.
\medskip\\
{\bf E2.} The maximal energy of photoelectrons 
is given by
\be\la{maxen}
\fr{\cm v_{\rm max}^2}2=\h\om-W~,
\ee
where $W$ is the {\it work function} of the substance.
Hence, the 
emission of the electron is possible only if
$\h\om-W>0$; thus, the redbound 
$\om_{\rm red}=W/\h$ does not depend on the  intensity of 
the  light --
this agrees with Lenard's observations !
\medskip\\
{\bf E3.}
Respectively, 
the stopping voltage should satisfy the inequality
\be\la{Vs}
-eU_{\rm stop}>\h\om-W~,
\ee
where $e<0$. Thus the minimal $U_{\rm stop}$ also does not 
depend on the  intensity of 
light.
\medskip

Formula  (\re{maxen}) formally represents the 
energy conservation in the absorption of the 
photon by the electron. 
However, let us stress that (\re{maxen})
is a theoretical interpretation of 
formula 
(\re{Vs}), which is verified experimentally and gives the 
minimal stopping voltage $-(\h\om-W)/e$. Moreover,
formula (\re{Vs}) allows to measure the Planck constant
$\h$
with high precision.


Thus the `Einstein rules' 
{\bf E1} -- {\bf E3}
give the complete expanation for  
Lenard's observations.
In 1922 A. Einstein was awarded the 1921 Nobel Prize 
in Physics for his 
theory of the 
photoelectric effect,
relying on the revolutionary {\it corpuscular
theory of light}.
\medskip

In 1927  G. Wentzel calculated angular 
distribution of the photocurrent
applying the first order
perturbation approach to the coupled Maxwell-
Schr\"odinger equations. 
\medskip

We give a slightly formalized version of 
Wentzel's calculations 
\cite[Vol. II]{Som}.
Namely, we justify
Wentzel's calculations and
 Einstein's rules 
by the limiting amplitude principle
in the framework of the perturbation approach.
We show that 
the photoeffect is caused by  slow decay of
the  limiting amplitude
 at infinity for  $|\om|>|\om_1|$.
The slow decay 
results in a
nonvanishing current to infinity; this means the 
photoelectric effect. Thus, $\om_{\rm red}=|\om_1|$.
Moreover, 
the photoelectron energy is given by (\re{maxen}), and
stopping voltage satisfies (\re{Vs}).

Unfortunately, the perturbation approach is not selfconsistent, and 
should be considered, rather as a hint for explaning  the 
atomic ionization. The corresponding rigouros theory of    
ionization was developed recently \ci{Cos01}-\ci{CLST2010}.
However, the theory implies the atomic ionization for 
any light frequency $\om\ne 0$.
For second quantized models 
a perturbation treatment of 
the atomic ionization and of relation 
 (\re{maxen})
 were given in 
\ci{Bach01, GZ2009, Z2006}.

This being so, a dynamical nonperturbation
explanation of Einsten's rules for
photoelectric effect remains an open challenging problem.

\setcounter{subsection}{0}
\setcounter{theorem}{0}
\setcounter{equation}{0}
\section{Scattering problem}
We want to describe the scattering of light by the Hydrogen 
atom in its ground state. 
The scattering
is described 
by the coupled Maxwell-Schr\"odinger equations in the
Born approximation
\be\la{SMes}
\ds[i\h\pa_t-e\phi_n(\x)]\psi(t,\x) =\fr 1{2\cm}
[-i\h\nab-\ds\fr ec \bA_0(t,\x)]^2\psi(t,\x) 
\ee 
\be\la{SMeas}
\left\{\ba{l}
\ds\fr 1{4\pi}\Box \phi(t,\x)=\rho(t,\x)=e|\psi(t,\x)|^2~,\\
\\
\ds\fr 1{4\pi}\Box \bA(t,\x)=\ds\fr{\bj(t,\x)}c =\ds\fr e {\cm
c}[-i\h\nab-\ds\fr ec  \bA_0(t,\x)]\psi(t,\x ) \cdot\psi(t,\x) \ea
\right.~,
\ee
where $\phi_n(\x)=-e/|\x|$ is the Coulomb potential of
the nucleus, and $\cdot$ stands for the real scalar product 
of the complex numbers considered as vectors from $\R^2$:
$z_1\cdot z_2=\rRe z_1\ov z_2$. 
Further,
 $\bA_0$ stands for the incident  wave
 \be\la{A1t} ~~~~\bA_0(t,\x)=A\Theta(ct-x^1)\sin
k(x^1-ct)(0,0,1)~. 
\ee
where $\Theta$ is the Heaviside function, and
 $k$ is a wave number.
The incident wave is a  solution
to the Maxwell equations in {\it free space}:
\be\la{dps}
~~~~\Box \bA_0(t,\x)=0~,~~~~~(t,\x)\in\R^4~.
\ee

In
our model (\re{SMes}), (\re{SMeas}) the hydrogen
nucleus is considered as fixed. This corresponds to 
the fact that
nucleus is heavy with respect to the electron.

The hydrogen ground state energy is 
$E_1=-2\pi\h cR =-\cm e^4/(2\h^2)$, and
the corresponding  wave function  is
$\psi_1(\x)=C_1e^{-|\x|/r_1}$ (we assume that the atom is situated
at the origin). Then the corresponding solution to the Schr\"odinger
equation is 
\be\la{ceig}
\psi_1(t,\x)=C_1e^{-|\x|/r_1}e^{-i\om_1t}~,\,\,\,\,\om_1=\ds\fr{E_1}\h
=-\ds\fr{\cm e^4}{2\h^3}~,~~~~~r_1=\ds\fr{\h^2}{me^2}~.
\ee 

\setcounter{subsection}{0}
\setcounter{theorem}{0}
\setcounter{equation}{0}

\section{First order approximation}
We apply the 
perturbation approach
 expanding the
solution of  (\re{SMes}) for small
amplitudes $|A|$: 
\be\la{fa}
\psi(t,\x)=\psi_1(t,\x)+Aw(t,\x)+\cO(A^2)~,~~~~~~|A|\ll 1 
\ee 
where
$\psi_1(t,\x)$ is the groundstate (\re{ceig}). 
We suppose that the atom is in its groundstate for $t<0$, i.e.,
\begin{equation}\la{faw}
w(t,{\mathbf{x}})=0,~~~~~~~~t<0.
\end{equation}
Substituting (\re{fa}) into
(\re{SMes}), we obtain, in the first order in $A$, 
\be\la{SMes1}
A\ds[i\h\pa_t-e\phi_n(\x)]w(t,\x) =A\fr 1{2\cm} [-i\h\nab]^2
w(t,\x)+\fr{i\h e}{\cm c} \bA_0(t,\x)\cdot\nab \psi_1(t,\x) 
\ee
since $\psi_1(t,\x)$ is a solution to equation (\re{SMes}) with
$\bA_0=0$. 
By (\re{A1t})  and  (\re{ceig}), we have the following
splitting for the
source term in the RHS of (\re{SMes1}),
\beqn\la{SMes2} 
\fr{i\h e}{\cm c} \bA_0(t,\x)\cdot\nab
\psi_1(t,\x)&=& \fr{i\h e}{\cm c} A\sin k(x^1-ct)(0,0,1)\cdot\nab
\psi_1(\x) e^{-i\om_1 t}\nonumber\\
&=&
\fr{A\h e}{2\cm c}  [e^{i k(x^1-ct)}-e^{-i k(x^1-ct)}]e^{-i\om_1 t}
\nab_3\psi_1(\x)\nonumber\\
&=&\psi_+(\x)e^{-i(\om_1+\om)t}-\psi_-(\x)e^{-i(\om_1-\om)t}~,~~~t>x^1-c~,
\eeqn
where $\om:=kc$.
Now let us apply the {\it limiting amplitude principle}:
\be\la{lapra}
~~~~~~~w(t,\x)= w_+(\x)e^{-i(\om_1+\om)t}-w_-(\x)e^{-i(\om_1-\om)t}
+\sum_l C_l\psi_l(\x)e^{-i\om_l t}+r(t,\x)
\ee
where $w_\pm(\x)$ are the {\it limiting amplitudes},
and
$r(t,\cdot)\to 0$ as $t\to\infty$, in an appropriate norm.
Here
$\psi_l(\x)$ denote 
eigenfunctions of the discrete
spectrum of homogeneous Schr\"odinger equation
\be\la{hS}
\ds[i\h\pa_t-e\phi_n(\x)]w(t,\x)
=\fr 1{2\cm}
[-i\h\nab]^2 w(t,\x)~.
\ee 
The asymptotics (\re{lapra}) hold 
provided
 $\om_1\pm\om\ne\om_l$ for all $l$.

The sum over the discrete spectrum on the RHS
of (\re{lapra}) vanishes 
by (\re{faw}). (In any case,  
the sum  does not contribute to the photocurrent
since the eigenfuctions  rapidly decay  
at infinity.)
Then, in the first order approximation,
\be\la{was}
w(t,\x)\sim w_+(\x)e^{-i(\om_1+\om)t}-w_-(\x)e^{-i(\om_1-\om)t}~,
\,\,\,\,t\to+\infty
\ee
For $w_\pm$, we get equations
\be\la{wequ}
\ds[\om_1\pm \om+\ds\fr{e^2}{\h|\x|}]w_\pm(\x)+
\fr \h{2\cm}
\De w_\pm(\x)=\fr{\psi_\pm(\x)}{\h A}=\fr{e}{2\cm c}e^{\pm i kx^1}
\nab_3\psi_1(\x)~.
\ee

\setcounter{subsection}{0}
\setcounter{theorem}{0}
\setcounter{equation}{0}

\section{Radiation in continuous spectrum}
We will consider scattering of light with large frequencies:
\be\la{fbn}~~~~~~~~~~~~~~~~|\om|>|\om_1|~.
~~~~~~~~~~~~~~~~~~~~~~~~~~~~~~~~~
\ee
For  the
Hydrogen we have $|\om_1|=\ds\fr{\cm e^4}{2\h^3}
\approx 20,5\cdot 10^{15} ~s^{-1}$ by (\re{ceig}). Hence, bound
(\re{fbn}) holds for wave numbers $|k|>k_1:=|\om_1|/c\approx 68\cdot 10^{7}$ m$^{-1}$
or wave lengths
$\lam < 2\pi/k_1=0.91176\cdot 10^{-5}$ cm $=911.76 \stackrel\circ A$.

For simplicity of notations, we assume that
$\om>0$.
Then $\om_1-\om<0$, but 
$$
\om_1+\om=\om-|\om_1|>0
$$
by (\re{fbn}).
Hence, $\h(\om_1+\om)$ belongs to the continuous spectrum
of the  stationary Schr\"odinger equation (\re{wequ}).
Therefore,
the solution $w_+(\x)\not\in L^2$. This means a
slow decay of the limiting amplitude: 
$$
|w_+({\mathbf{x}})|\sim 
\frac
{a(\n({\mathbf{x}}))}
{|{\mathbf{x}}|}~,
~~~~~~~~~
|{\mathbf{x}}|\to\infty~,
$$
where $\n({\mathbf{x}}):={\mathbf{x}}/|{\mathbf{x}}|$. 
We will calculate the 
amplitude
$a(\n)$
and obtain the 
main term of the radiation in the form
\be\la{mte}
Aw_+(\x)e^{-i(\om_1+\om)t}\sim
A
\fr{a(\vp,\theta)}{|\x|}e^{i[k_r|\x|-(\om+\om_1)t]}~,\,
\,\,\,|\x|\to\infty~.
\ee
On the other hand, $\h(\om_1-\om)<0$ does not belong to the
continuous spectrum of the Schr\"odinger equation.
Hence, $w_-(x)$ exponentially decays,
\be\la{wmL}
|w_-(\x)|\le Ce^{-\ve_-|\x|}~,~~~~~~~~\x\in\R~,
\ee
where $\ve_->0$.
In fact,
we can neglect 
term with $\ds\fr{e^2}{\h|\x|}$
in equation (\re{wequ}), since it is relatively small and decays at infinity. Then we obtain
\be\la{weqt}
[\De+\ds z_-] w_-(\x)=f_-(\x)~,\,\,\,\,\x\in\R^3
\ee
where $z_-={2m}(\om_1- \om)/\h<0$ and
$|f_-(\x)|\le Ce^{-\ve|\x|}$ with $\ve=-1/r_1>0$.
Hence,
$w_-=E_-*f_-$, where $E_-(\x)$ is the
fundamental solution $E_-(\x)=-e^{-\kappa_-|\x|}/(4\pi|\x|)$ with
$\kappa_-:=\sqrt{-z_-}>0$:
\be\la{cfs1}
w_-(\x)=-\int \fr{e^{-\ka_-|\x-\y|}}
{4\pi|\x-\y|}f_-(\y)d\y~.
\ee
As a result, 
decay (\re{wmL}) holds with
$\ve_-=\min(\ve, \kappa_-)>0$.
\medskip

We will deduce from (\re{mte}) and (\re{wmL})
the following asymptotics for the limiting stationary
electric current at infinity,
\be\la{elc}
\bj(t,\x)\sim A^2\fr{e\h k_r}{\cm}
\fr{a^2(\vp,\theta)}{|\x|^2}\n(\x)~,\,\,\,\,|\x|\to\infty~.
\ee
The formula
was obtained by Wentzel in 1927  (see \ci{We}) with amplitude
\be\la{amp}
a(\vp, \theta)=C\sin\theta\cos\vp~,
\ee
where $C\ne 0$.
Hence, 
formula (\re{elc})
describes a non-zero electric current from the
atom to infinity.
Indeed,
asymptotics (\re{elc}) 
imply that {\it total
electric current to infinity} does not vanish, i.e.,
\be\la{teci}
J_\infty:=\lim_{R\to\infty}\int_{|\x|=R}\bj(t,\x)dS(\x)\ne 0~.
\ee

\setcounter{subsection}{0}
\setcounter{theorem}{0}
\setcounter{equation}{0}

\section{The limiting amplitude}

Let us calculate the limiting amplitude $w_+(\x)$.
First, we rewrite equation (\re{wequ})
as follows,
\be\la{weqf}
[\nab^2 +k_r^2(\om)]w_+(\x)=
\fr{e}{\h c}e^{i kx^1}\nab_3\psi_1(\x)-\fr{2e^2\cm}{\h^2|\x|}w_+(\x)~,
\ee
where $k_r(\om):=\sqrt{2\cm(\om_1+\om)/\h}>0$.
In the first approximation, 
we can neglect 
the last term on the  RHS, because 
it is small and decays at infinity.
Then we get the Helmholtz equation
\be\la{weqg}
[\De +k_r^2(\om)]w_+(\x)=f_+(\x):=
\fr{ e}{\h c}e^{ i kx^1}\nab_3\psi_1(\x)~.
\ee
Hence the exponential decay (\re{wmL}) does not hold for 
$w_+(\x)$. This is obvious in the Fourier space where (\re{weqg})
becomes
\be\la{weqg2}
\hat w_+(\bk)=\ds\fr{\hat f_+(\bk)}{-\bk^2+k_r^2(\om)]}~.
\ee
The denominator vanishes on the sphere $|\bk|=k_r(\om)$,
while $\hat f_+(\bk)\sim \bk_3\hat\psi_1(\bk_1+k,\bk_2,\bk_3)$ 
is zero only for $\bk_3=0$.
Hence
$w_+(\x)$ cannot decay exponentially. Now
 the solution is given by the convolution 
\be\la{cfs}
w_+(\x)=-\int \fr{e^{ik_r(\om)|\x-\y|}}{4\pi|\x-\y|}f_+(\y)d\y~.
\ee
This follows from the 
limiting absorption principle 
since
fundamental solution
$\ds\fr{e^{ik_r(\om+i\ve)|\x-\y|}}{4\pi|\x-\y|}$
is a tempered distribution for small $\ve>0$, since
  $\rIm k_r(\om+i\ve)>0$ for the fixed branch
$k_r(\om)>0$.

Now we can calculate asymptotics (\re{mte}).
To do so we substitute expression
(\re{weqg}) for $f_+$ into (\re{cfs}). 
By partial integration,
we obtain
\beqn\la{cfp}
w_+(\x)&=&-\fr{e}{\h c}\int
\nab_{\y_3}\fr{e^{ik_r|\x-\y|}}{4\pi|\x-\y|}e^{ik\y^1}\psi_1(\y)d\y
\nonumber\\
\nonumber\\
&=&\fr{ik_r e}{\h c}\int
\fr{e^{ik_r|\x-\y|}(\x^3-\y^3)}{4\pi|\x-\y|^2}e^{ik\y^1}\psi_1(\y)d\y
+\cO(|\x-\y|^{-2})~.
\eeqn
Recall that $\theta$ denotes the 
angle between $\n:=\x/|\x|$ and $\e_1$, and 
$\vp$ stands for {\it azimuthal} angle between $\e_3$ and 
the
plane $(\n,\e_1)$. Then
$\x^3=\sin\theta\cos\vp|\x|$. Hence, 
(\re{cfp})
implies
(\re{mte}) with angular distribution
(\re{amp}),
because the ground state $\psi_1(\y)$ decays rapidly 
 at infinity.
The constant $C$ in (\re{amp}) is given by
\be\la{cC}
C=C(k)=\fr{ik_r e}{4\pi \h c}\int
e^{ik\y^1}\psi_1(\y)d\y~.
\ee
It does not vanish for the groundstate (\re{ceig}).

\setcounter{subsection}{0}
\setcounter{theorem}{0}
\setcounter{equation}{0}
\section{Angular distribution of  photocurrent:  The Wentzel formula}
Our aim is to derive  (\re{elc}).
Equations (\re{SMeas}) imply that,
in the first approximation, 
the current is given by
\be\la{SMec}
\bj
:=-\ds\fr e \cm[i\h\nab\psi(t,\x )]
\cdot\psi(t,\x )~.
\ee
Further, $w_-(\x)$ decays exponentially at infinity by
(\re{wmL}), as well as  the eigenfunctions 
of the discrete spectrum
$\psi_l(\x)$. 
Therefore, (\re{fa}) 
and (\re{was})
imply the asymptotics
\begin{equation}\la{fab}
\psi(t,\x)\sim
Aw_+(\x)e^{-i(\omega_1+\omega)t}~,\qquad
|\x|\to\infty.
\end{equation}
Substituting
into (\re{SMec}), and using
asymptotics (\re{mte}), we obtain the Wentzel formula (\re{elc})
with amplitude (\re{amp}).

\setcounter{subsection}{0}
\setcounter{theorem}{0}
\setcounter{equation}{0}

\section{Derivation of Einstein's rules}

Now we can explain Lenard's observations and Einstein's rules for
the photoelectric effect:
\bigskip\\
{\bf E1} By (\re{elc}), the saturation photocurrent is
proportional to $A^2$, which in turn is proportional to the
intensity of incident light.
\medskip\\
{\bf E2} Asymptotics (\re{mte}) imply that
the energy per one photoelectron is given by the {\it Einstein
formula} (\re{maxen}). Indeed, for large $|{\mathbf{x}}|$, the
radiated wave (\re{mte}) is locally close to the plane wave
with the frequency $\omega-|\omega_1|$. Hence,
the energy per one photoelectron
is given by $E=\hbar(\omega-|\omega_1|)$, which is equivalent to
(\re{maxen}) with 
\be\la{Who}
W=\hbar\omega_1~.
\ee
{\bf E3} Application of stopping voltage is equivalent to the
corresponding modification of the scalar potential in the
Schr\"odinger equation (\re{SMes}): $\phi_n({\mathbf{x}})\mapsto
\ti\phi({\mathbf{x}})=\phi_n({\mathbf{x}})+\phi_{\rm
stop}({\mathbf{x}})$, where $\phi_{\rm stop}({\mathbf{x}})$ is
a~slowly varying potential, and $\phi_{\rm
stop}({\mathbf{x}})=U_{\rm stop}>0$ in a~macroscopic region
which 
contains the atom. 
Therefore,  the ground state energy
$\hbar\omega_1$ changes to $\hbar\ti\omega_1$, and 
$\hbar\ti\omega_1\approx\hbar
\omega_1+eU_{\rm stop}$ with high precision.
Indeed, by the
 Courant minimax principle,
\begin{equation}\la{HC}
\hbar\omega_1=\min\limits_{\Vert\psi\Vert=1} (\psi,H\psi)~.
\end{equation}
We can assume that $0\le \phi_{\rm stop}({\mathbf{x}})\le U_{\rm
stop}$ for ${\mathbf{x}}\in{\mathbb R}^3$. Then
\begin{equation}\la{HC2}
\hbar\ti\omega_1=\min\limits_{\Vert\psi\Vert=1}
(\psi,[H+e\phi_{\rm stop}({\mathbf{x}})]\psi)\ge
\hbar\omega_1+eU_{\rm stop}~.
\end{equation}
On the other hand, the unperturbed ground state 
$\psi_1({\mathbf{x}})$ is localized in a very small region of the
size about $1\st{0}A=10^{-8}$ cm, where $e\phi_{\rm
stop}({\mathbf{x}})=U_{\rm stop}$. Hence,
\begin{equation}\la{HC3}
(\psi_1,[H+e\phi_{\rm stop}({\mathbf{x}})]\psi_1)\approx
\hbar\omega_1+eU_{\rm stop}~.
\end{equation}
Therefore, 
\begin{equation}\la{HC22}
\hbar\tilde\omega_1\approx
\hbar\omega_1+eU_{\rm stop}~.
\end{equation}
For the eigenstates with highest numbers, the localization gets
progressively worse, and the eigenfunctions of the continuous
spectrum are not localized at all.
Respectively, the shift of the highest eigenvalues is smaller and
smaller, and the continuous spectrum of the modified Schr\"odinger
operator remains unchanged.

Finally, the potential 
prevents the photoelectric effect if
the spectral condition (\re{fbn}) fails for the modified ground
state; i.e., $0<\omega<|\ti\omega_1|$ or
\begin{equation}\la{fbn2}
\hbar\omega<|\hbar\omega_1+eU_{\rm stop}|=\hbar|\omega_1|-eU_{\rm
stop}~,
\end{equation}
since $\omega_1<0$, while $e<0$ and we define 
$U_{\rm stop}>0$. In other words,
\begin{equation}\la{fbn3}
-eU_{\rm stop}>\hbar\omega-\hbar|\omega_1|~,
\end{equation}
which is equivalent to (\re{Vs}) by (\re{Who}).

\setcounter{subsection}{0}
\setcounter{theorem}{0}
\setcounter{equation}{0}

\section{Further improvements}
The Wentzel calculation takes into account  
interaction of the Maxwell and Schr\"odinger fields
in the first order of approximation. Next,
second order correction,
was obtained by Sommerfeld
and Shur \ci{SS}.
 The corresponding
corrected formula reads (see \ci[Vol. II]{Som})
\be\la{elcnSS}
\bj(t,\x)\sim
\fr{\sin^2\theta\cos^2\vp(1+4\beta\cos\theta)}{|\x|^2}
~\n(\x)~,\,\,\,\,|\x|\to\infty~.
\ee
Here $\beta=\ds\fr vc$, where $v$ is velocity of the
photoelectrons. The formula means an {\it increment}
 of the scattering amplitude for angles $0<\theta<\ds\fr\pi 2$
and a {\it decrement} of the scattering amplitude for
angles $\ds\fr\pi 2<\theta<\pi $. This means a
{\it forward shift of scattering} due to
{\it pressure of the
incident light upon the outgoing photocurrent},
as predicted by Wentzel
\ci{We}.

Fisher and Sauter have obtained
the formula which holds in each order
(see \ci[Vol. II]{Som}):
\be\la{elcnSSao}
\bj(t,\x)\sim
\fr{\sin^2\theta\cos^2\vp}{(1-\beta\cos\theta)^4|\x|^2}
\n(\x)~,\,\,\,\,|\x|\to\infty~.
\ee

\setcounter{subsection}{0}
\setcounter{theorem}{0}
\setcounter{equation}{0}

\section{Ionization and photoeffect}
Unfortunately, the perturbation theory 
of the photoeffect
is not 
selfconsistent. For instance, 
the stationary nonvanishing photocurrent  (\re{teci})
contradicts the charge conservation
law, because
the atomic charge is finite.
The contradiction is provided by the perturbation strategy,
 which leaves  unchanged $\psi_1$
on the right hand side of  (\re{SMes1})
while it should be substituted by the solution
of (\re{SMes}), (\re{SMeas}).
This `selfaction' should result in a decay of the photocurrent
until the negative atomic charge will be exausted, i.e.,
\be\la{ioniz}
\int_{|\x|<R}|\psi(t,\x)|^2d\x\to 0~,~~~~~~~t\to\infty
\ee
for all $R>0$.
Thus, formula (\re{teci}) does not justify the photoeffect
but it rather suggests the {\it atomic  ionization} (\re{ioniz}),
as
established in \ci{Cos01}-\ci{CLST2010}.
\medskip\\
On the other hand, a 
selfconsistent justification of the  photoeffect
should rely on the stationary photocurrent, since 
the stopping voltage is concerned exactly 
the stationary picture.
To maintain the stationary photocurrent, one needs either 
an external
source (galvanic element, etc) to reimburse the charge
decay, or 
a different model with infinite charge (e.g., crystal). 
\medskip\\
Finally, the ionization occurs at any light frequency $\om$
different from zero, according to the results 
of \ci{Cos01}-\ci{CLST2010}.
Thus, a satisfactory nonperturbation explanation of 
the Einstein rules of the photoelectric effect 
remains still an open problem.



\begin{thebibliography}{99}




\bibitem{Bach01}
V. Bach, F. Klopp, H. Zenk,
 Mathematical analysis of the photoelectric effect
{\em Adv. Theor. Math. Phys.} {\bf  5} (2001), 969-999. 
arXiv:math-ph/0210051


\bibitem{Cos01}
O. Costin, R.D. Costin, J.L. Lebowitz, A. Rokhlenko,
Evolution of a model quantum system under time periodic forcing:
conditions for complete ionization,
{\em Commun. Math. Phys.} {\em 221} (2001), no.1, 1-26.


\bibitem{Cos04}
O. Costin, R.D. Costin, J.L. Lebowitz,
Time asymptotics of the Schr\"odinger wave
function in time-periodic potentials,
{\em J. Stat. Phys.} {\bf 116} (2004), no. 1-4, 283-310.



\bibitem{Cos08}
O. Costin, J.L. Lebowitz, C. Stucchio,
Ionization in a 1-dimensional dipole model,
{\em Rev. Math. Phys.} {\bf 20}, no. 7 (2008), 835-872.


\bibitem{CLST2010}
O. Costin, J.L. Lebowitz, C. Stucchio, S. Tanveer,           
Exact results for ionization of model atomic systems,
{\em J. Math. Phys.} {\bf 51} (2010), no. 1, Paper No. 015211, 16 p.
                              



\bibitem{GZ2009}
M. Griesemer, H. Zenk,
On the Atomic Photoeffect in Non-relativistic QED,
arXiv:0910.1809



\bibitem{Som}
A. Sommerfeld,
Atombau und Spektrallinien, Vol. I and II, Friedr. Vieweg $\&$
Sohn, Brounschweig, 1951.



\bibitem{SS}
A. Sommerfeld, G. Schur, {\em Ann. d. Phys.} {\bf 4} (1930), 409.


\bibitem{We}
G. Wentzel, {\em ZS. f. Phys.} {\bf 43} (1927), 1, 779.



\bibitem{Z2006}
H. Zenk,
Ionisation by quantised electromagnetic fields: 
The photoelectric effect, 
{\em Rev.Math.Phys.} {\bf 20} (2008), 367-406.





\end{thebibliography}
\end{document}